\documentclass[prl,aps,twocolumn,showpacs,amsmath,amssymb]{revtex4-1}

\usepackage{graphicx}
\usepackage{psfrag}
\usepackage{epsfig}
\usepackage{color}
\usepackage[tight]{subfigure}
\definecolor{dred}{rgb}{0.7,0.0,0.0}
\usepackage{dcolumn}
\usepackage{bm}
\allowdisplaybreaks 
\usepackage{braket}
\usepackage{ulem} 
\usepackage{hyperref}
\hypersetup{colorlinks=true, linkcolor=blue, citecolor=blue, filecolor=blue, urlcolor=blue, pdftitle=, pdfauthor=, pdfsubject=, pdfkeywords=}

\definecolor{orange}{rgb}{1,0.5,0}
\definecolor{black}{rgb}{0,0,0}

\begin{document}

\title{Fractional Chern insulators with strong interactions 
far exceeding bandgaps}

\author{Stefanos Kourtis}
\affiliation{Institute for Theoretical Solid State Physics, IFW Dresden, 01171 Dresden, Germany}

\author{Titus Neupert}
\affiliation{Princeton Center for Theoretical Science, Princeton University, Princeton,
New Jersey 08544, USA}

\author{Claudio Chamon} \affiliation{ Physics Department, Boston
  University, Boston, Massachusetts 02215, USA }
            
\author{Christopher Mudry} \affiliation{ Condensed Matter Theory
  Group, Paul Scherrer Institute, CH-5232 Villigen PSI, Switzerland}

\date{\today}

\begin{abstract}
  We study two models for spinless fermions featuring
  topologically non-trivial bands characterized by Chern numbers
  $C=\pm1$ at fractional filling. Using exact
  diagonalization, we show that, even for infinitely strong
  nearest-neighbor repulsion, the ground states of these models belong
  to the recently discovered class of quantum liquids called
  fractional Chern insulators (FCI). We thus establish that
  FCI states can arise even if interaction strengths are arbitrarily 
  larger than the noninteracting band gap, going beyond
  the limits in which FCI states have been previously studied.  
   The strong-coupling FCI states therefore
  depart from the usual isolated-band picture that parallels the
  fractional quantum Hall effect in Landau levels and demonstrate
  how a topologically ordered state can arise in a truly multiband
  system.
\end{abstract}

\maketitle


The recently discovered prospect~\cite{Tang2011,Sun2011,Neupert2011} 
of realizing fractional quantum-Hall (FQH) states~\cite{Tsui1982} 
with the inclusion of short-range interactions in lattice models 
featuring fractionally filled, topologically non-trivial bands has 
garnered considerable interest. Apart from a paradigmatic extension 
of the FQH effect to lattices, these states, 
called fractional Chern insulators (FCI), 
arise without an externally applied magnetic field.
Hence, they are an important conceptual step towards technological applications. 
There are already several proposals for the realization of FCI states, 
some of which involve optical lattices~\cite{Yao2013a}, 
while others are based on known material structures, 
such as strained or irradiated graphene~\cite{Ghaemi2012a,Grushin2013}, 
oxide heterostructures~\cite{Xiao}, 
or layered multi-orbital systems~\cite{Venderbos2011a,Kourtis2012a}. 
The latter category is particularly promising, 
since the energy scales at which the desired physics emerges 
is of the order of room temperature.

Since their inception, FCI states have been studied extensively 
using various numerical and analytical methods%
~\cite{Parameswaran2013,Bergholtz2013} .
Of particular interest have been works that
emphasize the differences between FCI states and 
traditional FQH physics. Perhaps the most obvious difference, 
which was noticed early on, is that Chern bands, unlike Landau levels, 
have a non-vanishing dispersion. 
It was first proven by example in Ref.~\onlinecite{Kourtis2012a} 
and then substantiated more formally in Refs.%
~\onlinecite{Grushin2012,Chamon2012,Murthy2012,Lauchli2012} 
that this dispersion may actually favor FCI states. 
Unlike FQH systems, FCI models can be naturally extended 
to include both spin species, and it has been shown that 
the resulting time reversal-symmetric models can be hosts of 
fractional topological insulators~\cite{Neupert2011a}, 
a prospect that was envisaged prior to the advent of FCIs%
~\cite{PhysRevLett.96.106802,Levin2009}. Since the Chern number of a band can, 
in contrast to a Landau level, take values larger than one%
~\cite{Wang2011b,Trescher2012,Yang2012a}, 
FCI states can occur in partially filled bands 
with higher Chern numbers%
~\cite{Wang2011a,Grushin2012,Liu2012b,Sterdyniak2012}. 
Finally, topologically ordered states that go markedly beyond 
the Landau-level picture, in which the topological character 
is combined with Landau-type order, have been found recently%
~\cite{Kourtis2013}.

In the literature on FCIs, 
most works deal with a single isolated band. 
The presence of more than one bands has
been taken into account in few examples%
~\cite{Venderbos2011a,Kourtis2012a,Kourtis2013}, 
but the effect of band mixing has not been systematically studied. 
If one wishes to search for FCI states in the laboratory, 
understanding of how these states can arise in realistic multiband systems 
is crucial. In this manuscript, we wish to pose two fundamental questions 
for the realization of FCI states, namely (i) whether the mixing of bands by interactions 
leaves space for FCI states to arise, and (ii) whether FCI states can be found far beyond the energy scale of the band gap. 
We shall answer both questions positively.

We have studied two prototypical two-sublattice FCI models using exact
diagonalization, taking both Chern bands into account.  We show that
FCI states survive band mixing caused by arbitrarily large
interactions. To demonstrate this, we introduce the extreme limit of
nearest-neighbor interaction going to infinity.  In this regime, which
can be exploited further in the study of higher dimensional and
spinful systems, particles dressed by the interaction form extended
objects, which can be interpreted as non-interacting hardcore particles
occupying more than one lattice sites. We find that strong
interactions of magnitude far larger than the band gap may actually
favor FCI states, regardless of whether the bands are mixed.
These observations provide fresh insights into topological
ordering, and more importantly, they open prospects to realizing
experimentally fractional topological states of matter at high
temperatures.


\paragraph{Models} ---
We consider two models for interacting
spinless fermions that hop on a lattice
with two inequivalent sites per unit cell.
We shall endow both models
with topologically non-trivial band structures. 
They have the general form
\begin{subequations}
\begin{equation}
\hat{H}:= 
\hat{H}^{\,}_{\textrm{kin}} 
+ 
\hat{H}^{\,}_{\textrm{int}}\,.
\end{equation}
The kinetic energy
$\hat{H}^{\,}_{\textrm{kin}}$ is 
\begin{equation}
\hat{H}^{\,}_{\textrm{kin}}:= 
\sum_{{\bm{k}}\in\textrm{BZ}} 
\hat{\psi}^{\dag}_{{\bm{k}}}\, 
\mathcal{H}^{\,}_{\bm{k}}\, 
\hat{\psi}^{\,}_{\bm{k}}\,,
\end{equation}
where 
$\hat{\psi}^{\dag}_{{\bm{k}}}\equiv 
\left(\hat{c}^{\dag}_{{\bm{k}},A}\,,\hat{c}^{\dag}_{{\bm{k}},B}\right)$ 
denotes an operator-valued spinor
whose upper and lower components create spinless fermions 
with the wave-number $\bm{k}$ from the Brillouin zone (BZ)
on the inequivalent sites $A$ and $B$, respectively. 
The $2\times2$ matrix $\mathcal{H}^{\,}_{{\bm{k}}}$ is
\begin{equation}
\mathcal{H}^{\,}_{{\bm{k}}}:= 
g^{\,}_{0,\bm{k}}\,
\tau^{\,}_{0}
+ 
\bm{g}^{\,}_{\bm{k}}
\cdot
\bm{\tau} 
+ 
\mu^{\,}_{\mathrm{s}}\, 
\tau^{\,}_{3}\,,
\end{equation}
where we have introduced the $2\times2$ unit matrix $\tau^{\,}_{0}$
together with the three Pauli matrices 
$\bm{\tau}=(\tau^{\,}_{1},\tau^{\,}_{2},\tau^{\,}_{3})$ 
acting on the indices $A$ and $B$. 
The functions 
$g^{\,}_{0,\bm{k}}$,
$g^{\,}_{1,\bm{k}}$,
$g^{\,}_{2,\bm{k}}$,
and
$g^{\,}_{3,\bm{k}}$
are smooth real-valued functions of the wave-number $\bm{k}$ in the
thermodynamic limit
and we have made explicit the dependence on the staggered
chemical potential $\mu^{\,}_{\mathrm{s}}\in\mathbb{R}$.
We shall study two specific examples below,
by specifying the functions 
$g^{\,}_{\mu,\bm{k}}$ with $\mu=0,1,2,3$.
The interaction term $\hat{H}^{\,}_{\textrm{int}}$ 
is the nearest-neighbor repulsion
\begin{equation}
\hat{H}^{\,}_{\textrm{int}}:= 
V 
\sum_{\langle\bm{i},\bm{j}\rangle} 
\hat{n}^{\,}_{\bm{i}}\, 
\hat{n}^{\,}_{\bm{j}}, 
\end{equation}
\end{subequations}
where $V\geq0$ is the strength of the nearest-neighbor repulsion,
$\langle\bm{i},\bm{j}\rangle$ are directed nearest-neighbor bonds, 
and $\hat{n}^{\,}_{\bm{i}}$ 
is the number operator that counts how many spinless fermions
occupy the lattice site 
$\bm{i}\in\Lambda\equiv\Lambda^{\,}_{A}\cup\Lambda^{\,}_{B}$.


The checkerboard-lattice model of Refs.~\onlinecite{Neupert2011,Sun2011} 
can be written as
\begin{subequations}
\begin{align}
g^{\,}_{0,\bm{k}} &= 
4 t^{\,}_3 \cos k^{\,}_x \cos k^{\,}_y\,,
\\
g^{\,}_{1,\bm{k}} &= 
4 t \cos\varphi \cos \frac{k^{\,}_x}{2} \cos \frac{k^{\,}_y}{2}\,, 
\\
g^{\,}_{2,\bm{k}} &= 
4 t \sin\varphi \sin \frac{k^{\,}_x}{2} \sin \frac{k^{\,}_y}{2}\,, 
\\
g^{\,}_{3,\bm{k}} &= 
2 t^{\,}_2 ( \cos k^{\,}_x - \cos k^{\,}_y )\,,
\end{align}
where 
$t$, 
$t^{\,}_2$, 
and $t^{\,}_3$ are first nearest-, second nearest-, 
and third nearest-neighbor hopping amplitudes, respectively. 
In this definition, 
the primitive vectors of the checkerboard lattice have been chosen as 
$\bm{a}_1 = (\sqrt{2}/2,0)^{\mathsf{T}}$ 
and 
$\bm{a}_2 = (0,\sqrt{2}/2)^{\mathsf{T}}$, 
with the unit-cell sites being at points $(0,0)$ and $(1,1)$. 
In the following, we will fix $t^{\,}_2/t=0.4$ and $\varphi=\pi/4$. 
The flatness of the lower Chern band can be tuned by $t^{\,}_3$ 
and is maximized at $t^{\,}_3/t \approx 0.3$.
\label{eq:checkerboard}
\end{subequations}


The triangular-lattice model of Refs.~\onlinecite{Venderbos2011a,Kourtis2012a}
can be written as
\begin{subequations}
\begin{align}
g^{\,}_{0,\bm{k}}&= 
2t^{\,}_3 
\sum_{j=1}^3 
\cos(2{\bm{k}}\cdot\bm{a}^{\,}_j)\,,
\\
g^{\,}_{i,\bm{k}}&= 
2t \cos({\bm{k}}\cdot\bm{a}^{\,}_i), 
\qquad
i=1,2,3,
\end{align}
where 
$\bm{a}^{\,}_1 = (1/2, -\sqrt{3}/2)^{\mathsf{T}}$, 
$\bm{a}^{\,}_2 = (1/2, \sqrt{3}/2)^{\mathsf{T}}$, 
and $\bm{a}^{\,}_3 = -( \bm{a}^{\,}_1+\bm{a}^{\,}_2 )$ 
are the triangular-lattice unit vectors. 
The first nearest-neighbor and third nearest-neighbor hopping amplitude 
are $t$ and $t^{\,}_3$, respectively.
The third nearest-neighbor hopping amplitude
$t^{\,}_3$ can be used to tune the dispersion of the lower Chern band, 
with the flattest bands achieved for $t^{\,}_3/t \approx 0.2$.
\label{eq:triangular}
\end{subequations}


\paragraph{Infinite-$V$ limit}---
In the following, we will make use of the limit in which the
nearest-neighbor repulsive interaction strength $V$ is
taken to infinity.  In this case, particles cannot
occupy nearest-neighbor sites: any many-body state with
two spinless fermions sitting on neighboring sites is projected
out of the Hilbert space in this limit.

Thus, for any site $\bm{i}\in\Lambda$ 
we define the projected operator 
$\tilde{c}^{\dag}_{\bm{i}}$
by demanding that its action on any state in the occupation
basis of the projected Hilbert space is to
create a spinless fermion on $\bm{i}$ if and only if this site and
all its nearest-neighbor sites are empty. Otherwise,
$\tilde{c}^{\dag}_{\bm{i}}$ annihilates any state
from the  projected Hilbert space. Formally,
\begin{subequations}
\begin{equation}
\tilde{c}^{\dag}_{\bm{i}}:=
\hat{c}^{\dag}_{\bm{i}}
\prod_{\bm{j}\in\langle\bm{i}\bm{j}\rangle}
\left(1-\hat{n}^{\,}_{\bm{j}}\right),
\end{equation}
and
\begin{equation}
\begin{split}
\hat{H}=&\,
t 
\sum_{\langle\bm{i},\bm{j}\rangle} 
\left( 
e^{\mathrm{i}\phi^{\,}_{\bm{i},\bm{j}}} 
\tilde{c}^{\dag}_{\bm{i}}\, 
\tilde{c}^{\,}_{\bm{j}} 
+ \textrm{H.c.} 
\right)
\\
&\,
+ 
t^{\,}_2 
\sum_{\langle\langle\bm{i},\bm{j}\rangle\rangle} 
(-1)^{|\bm{i}|} 
\left( 
\tilde{c}^{\dag}_{\bm{i}}\,
\tilde{c}^{\,}_{\bm{j}} 
+ 
\textrm{H.c.} 
\right) 
\\
&\,
+ 
t^{\,}_3 
\sum_{\langle\langle\langle\bm{i},\bm{j}\rangle\rangle\rangle} 
\left( 
\tilde{c}^{\dag}_{\bm{i}}\, 
\tilde{c}^{\,}_{\bm{j}} 
+ 
\textrm{H.c.} 
\right) 
\\
&\,
+ 
\mu^{\,}_{\mathrm{s}} 
\sum_{\bm{i}} 
(-1)^{|\bm{i}|} \;
\tilde{c}^{\dag}_{\bm{i}}\,\tilde{c}^{\,}_{\bm{i}}
\,.
\end{split}
\end{equation}
\end{subequations}
Here,
$\phi^{\,}_{\bm{i},\bm{j}}$ 
are the phase factors needed to represent each of the two models
of 
Eqs.~\eqref{eq:checkerboard} 
and 
\eqref{eq:triangular} 
and we have set $|\bm{i}|$ to be even (odd) on sublattice $A$ ($B$). 
For the triangular-lattice model defined in
Eq.~\eqref{eq:triangular} with $t^{\ }_3=0$, 
this limit gives rise to the Hamiltonian
\begin{equation}
\hat{H}^{\,}_{\bigtriangleup}:= 
\sum_{\langle\bm{i},\bm{j}\rangle} 
\left( 
e^{\mathrm{i}\phi^{\,}_{\bm{i},\bm{j}}}\,
\tilde{c}^{\dag}_{\bm{i}}\, 
\tilde{c}^{\,}_{\bm{j}} 
+ 
\textrm{H.c.} 
\right)\,.
\end{equation}
Hamiltonian
$\hat{H}^{\,}_{\bigtriangleup}$
contains no free parameters.
$\hat{H}^{\,}_{\bigtriangleup}$ is similar to previously studied supersymmetric models~\cite{Fendley2003},
which yield exotic ``superfrustrated'' states with extensive groundstate degeneracy in many lattices~\cite{Fendley2005}.
Below we shall see that
$\hat{H}^{\,}_{\bigtriangleup}$
gives rise to FCI states at $\nu=1/3$ of the lower band, i.e., 1/6 filling of the full lattice.

The infinite-$V$ limit comes with a considerable reduction of 
the dimensionality of the Fock space~\cite{Zhang2003,Zhang2004}. Technically, 
this may be crucial in the search for new topological states, 
especially in higher dimensions,
where the lattice coordination and thus the reduction 
of the Hilbert space is typically higher. 
Evidently, taking further-neighbor repulsive interactions
$V^{\,}_{2}$,  $V^{\,}_{3}$, $\ldots$  
to the hardcore limit 
$V^{\,}_{2}\to\infty$,  $V^{\,}_{3}\to\infty$, $\ldots$  
allows for even more dramatic 
reductions of the Hilbert space.


\paragraph{Results \& discussion} ---
The general properties of the two models defined in
Eqs.~\eqref{eq:checkerboard} 
and 
\eqref{eq:triangular} 
have been presented elsewhere~\cite{Neupert2011,Kourtis2012a}. 
Here, we wish to highlight mainly two points, 
that may prove to be crucial in the search for FCI states: 
(i) Band mixing does not necessarily 
reduce the propensity to form FCI states. 
(ii) The strength of interactions can be much larger 
that previously thought of and can, in fact, be set to be infinite, 
without driving the system out of the FCI phase. 
(We find similar results for $\nu=1/5$.)

\begin{figure}
 \includegraphics[width=\columnwidth]{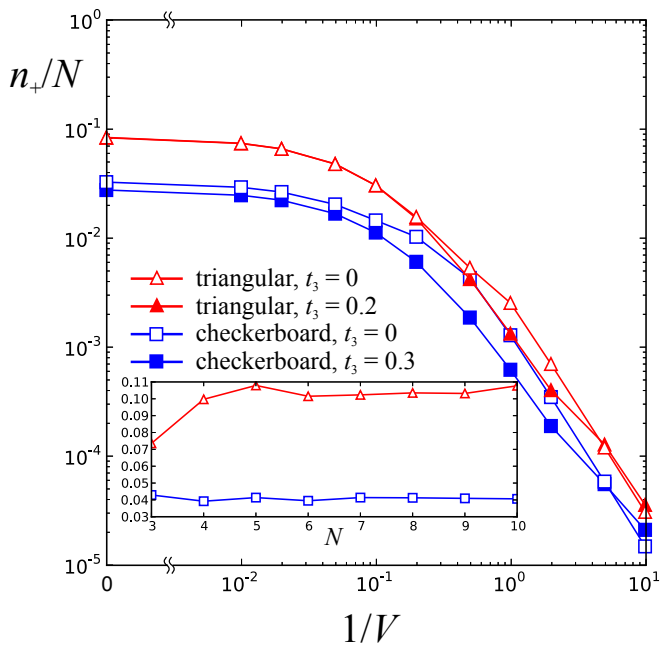}
\caption{(Color online) 
The density of upper-band character $n^{\,}_+$ for one of the
states in the FCI manifold
in the checkerboard (squares) and triangular (triangles) lattice models on a 
36-site cluster at $\nu=1/3$, with a kinetic term yielding almost flat (full symbols) 
and non-flat (empty symbols) lower bands, as a function of $1/V$. 
Inset: the value of $n^{\,}_+$ at $V=\infty$ 
as a function of system size, parametrized by the number of particles $N$.
        }
\label{fig:occupation} 
\end{figure}


We will now focus to the filling fraction $\nu=1/3$ 
of the lowest Chern band for each of the two models. 
We wish to know whether or how the inclusion of 
$\hat{H}_{\textrm{int}}$ in $\hat{H}$ 
will mix the bands formed by the eigenvalues of $\hat{H}_{\textrm{kin}}$. 
To this end, we have to measure the contribution -- if any -- 
of the upper band to the FCI states, that is, the overlap 
\begin{equation}
n^{\,}_+:=
\sum_{\bm{k}}
\left\langle E^{\,}_{0}\left|  
\hat{n}^{\,}_{\bm{k},+} 
\right|E^{\,}_{0}\right\rangle,
\end{equation} 
where $|E^{\,}_{0}\rangle$ 
is any one of the states in the degenerate ground-state manifold 
and $\hat{n}^{\,}_{{\bm{k}},+}$ is the operator measuring density of particles 
with upper-band character at wave-number $\bm{k}$~\cite{suppl}.
In Fig.~\ref{fig:occupation}, 
we show the expectation value of $n^{\,}_+$ for one of the
states in the FCI manifold
as a function of inverse interaction strength. 
We notice that, in the weak-coupling limit, 
band mixing is very limited. However, as the interaction reaches 
its maximal value, the mixing increases and saturates at appreciable values 
for both models. The contributions to the occupation
$n^{\,}_+$ are almost uniformly distributed across the Brillouin zone
in the finite clusters. Even though we cannot reach 
large enough system sizes for a finite-size extrapolation,
$n^{\,}_+$ shows no tendency of decreasing upon increasing system size,
as can be seen in the inset of Fig.~\ref{fig:occupation}.
The dependence of $n^{\,}_+$ on $V$ is almost
identical in the 48-site cluster with 8 particles and the 36-site cluster with 6 particles.

\begin{figure}
 \includegraphics[width=\columnwidth]{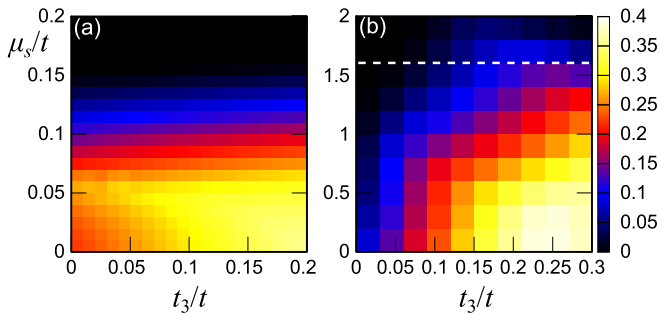}
\caption{(Color online) 
Phase diagrams of (a) the triangular and (b) 
the checkerboard lattice models on a 48-site cluster at 
$\nu=1/3,\, V=\infty$ in the $\mu^{\,}_{\mathrm{s}}$-$t^{\,}_3$ plane. 
The color coding is the lowest value of the gap between FCI ground states 
and excited states upon flux insertion~\cite{suppl}. 
The dashed white line in (b) denotes the phase boundary 
for the non-interacting model.
        }
\label{fig:phdiag}
\end{figure}

We will now show that, despite the fact that at $V=\infty$ the bands are mixed,
the FCI states remain robust for arbitrarily large interactions. 
To this end, we present the phase diagrams of the two models at $V=\infty$ 
in the $\mu^{\,}_{\mathrm{s}}$-$t^{\,}_3$ plane from Fig.~\ref{fig:phdiag}. 
We notice that, in both cases, the FCI phase is quite robust and 
does not depend crucially on the flatness of the original band. 
The FCI phase on the triangular lattice is quite sensitive 
to the introduction of the staggered chemical potential
$\mu^{\,}_{\mathrm{s}}$, 
presumably because this leads to an effective reduction in dimensionality 
at low energies. On the contrary, the FCI on the checkerboard lattice seems 
to be quite robust against 
$\mu^{\,}_{\mathrm{s}}$.  
(This phase seems to survive beyond the point where the bands 
of the non-interacting model would become topologically trivial,
but this may be a finite-size artifact.)

It should be mentioned that strong interactions may give rise to competing charge order~\cite{Kourtis2012a} 
or more exotic compositely ordered states~\cite{Kourtis2013},
whenever said competing orders are commensurate with the lattice. 
The results presented here are hence valid for short-range interactions 
at low enough densities, 
so that competing strong-coupling instabilities are ruled out.

\begin{figure}[t!!]
\includegraphics[width=\columnwidth]{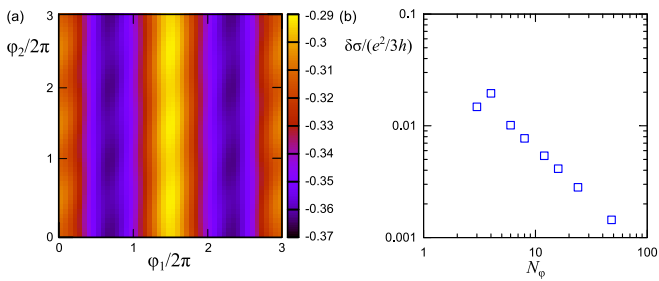}
\caption{(Color online) 
(a) Berry curvature and (b) relative error as a function 
of the square root of the number of points in the
flux-Brillouin zone partition for the triangular lattice model 
on a 48-site cluster at $\nu=1/3,\, V=\infty,\, t^{\,}_3 = 0$.
        }
\label{fig:berry}
\end{figure}

Within the FCI regime (the colored part of the phase diagrams), 
the ground-state eigenvalues exhibit the empirical characteristic features 
of FCI states: 3-fold degeneracy and spectral flow. In order to establish 
beyond doubt that the phase is indeed an FCI, however, 
we calculate the Hall conductivity
in this regime~\cite{suppl}. We find it to be 
very precisely quantized to the value $-1/3$. 
The Berry curvature, as well as the accuracy of the quantization 
are shown in Fig.~\ref{fig:berry}. 
We notice that the Berry curvature is 
a very smooth function of $\varphi^{\,}_1,\,\varphi^{\,}_2$.

\paragraph{Conclusions} ---
FCI states are an important recent addition 
to the arsenal of theoretically predicted, topologically non-trivial states. 
For these states to be ultimately useful, however, it is imperative 
to bring them closer to reality. In this manuscript we have presented 
results that do this in two ways. 
(i) FCI states are not limited to the energy scale of weak interactions, 
but can arise for arbitrarily strong repulsion, 
meaning that the search for such states can be extended to materials 
with strong correlations. (ii) FCI states are robust against 
appreciable band mixing and hence candidate systems need not have 
a very large gap to host them, 
even though the bands that are mixed have opposite Chern numbers,
in contrast to the case of band mixing in Landau levels~\cite{Sterdyniak2012a, Hafezi2007}.
We have based our conclusions 
on exact numerical evidence obtained with Lanczos diagonalization, 
that allow us to identify the character and the properties of 
the ground states on finite systems.

\begin{acknowledgments}
S.K. was supported by the Deutsche Forschungsgemeinschaft under the Emmy-Noether program 
(Grant DA 1235/1-1).
T.N. acknowledges financial support by the Swiss National Science Foundation.
C.C. was supported by DOE grant DEFG02-06ER46316.
\end{acknowledgments}

\bibliographystyle{apsrev4-1}
\bibliography{bib,notes}

\newpage

\

\newpage

\appendix
\begin{center}
 \textbf{APPENDIX}
\end{center}

\section{Characterization of FCI states}
\subsection{Properties of the many-body spectrum}
The universal spectral properties of Hamiltonians with FCI groundstates 
are analogous to those of FQH states~\cite{Tao1984,Wen1990}.
Laughlin-like FCI ground states on the torus are gapped, 
with a topological degeneracy equal to the denominator of the filling fraction 
$\nu$ of the partially filled Chern band. In finite systems, 
the ground-state degeneracy of FCI states is not exact. 
Due to this splitting, FCI ground-state eigenvalues exhibit spectral flow, 
meaning that they exchange values upon insertion of one flux quantum through 
one of the handles of the torus, 
whenever they reside in different momentum sectors. 
Flux insertion is defined as the transformation 
\begin{equation}
t^{\,}_{\bm{i},\bm{j}} 
\to 
t^{\,}_{\bm{i},\bm{j}}\,
e^{
\mathrm{i}
\left(
\varphi^{\,}_1 
\frac{j^{\,}_1-i^{\,}_1}{L^{\,}_1}
+ 
\varphi^{\,}_2 
\frac{j^{\,}_2-i^{\,}_2}{L^{\,}_2}
\right)
  }
\end{equation}
for the hopping from the site at position 
$\bm{i}\equiv(i^{\,}_1 \ i^{\,}_2)^{\mathsf{T}}$ 
to the site at position 
$\bm{j}\equiv(j^{\,}_1 \ j^{\,}_2)^{\mathsf{T}}$, 
where the components of the position vectors, 
as well as those of the flux vector 
$\boldsymbol\varphi\equiv(\varphi^{\,}_1 \ \varphi^{\,}_2)^{\mathsf{T}}$, 
are along the directions of the corresponding primitive lattice vectors, 
and $\mathrm{i}$ is the imaginary unit. 
Note that here we have chosen to distribute the flux phase equally to all hoppings, 
therefore dividing the fluxes $\varphi^{\,}_1$ and $\varphi^{\,}_2$ 
by the corresponding lattice extents $L^{\,}_1$ and $L^{\,}_2$, 
in order to maintain the translational invariance of the lattice. 
In every other respect, this choice is equivalent to twisting 
the boundary conditions. For the determination of the phase boundaries
in the phase diagrams of the main text, we consider the encountered 
ground states as gapped only if the corresponding energy levels 
do not cross excited-state levels upon flux insertion.

\subsection{Hall conductivity}
There are cases for which the
ground-state degeneracy and spectral flow may not be enough 
to distinguish a FCI from topologically trivial states. 
To unequivocally determine whether the ground states we obtain are FCI, 
we calculate the many-body ground-state Hall conductivity, 
defined as~\cite{Niu1985,Xiao2010}
\begin{equation}
\begin{split}
\sigma^{\,}_{\mathrm{H}}:=&\,
\frac{e^2}{h}
\frac{L^{\,}_1\,L^{\,}_2}{\pi q} 
\sum_{n=1}^{q} 
\int\limits_{0}^{2\pi} 
\mathrm{d}\varphi^{\,}_1 
\int\limits_{0}^{2\pi} 
\mathrm{d}\varphi^{\,}_2
\\
&\, 
\times
\mathrm{Im}\,
\sum_{n'\not=n} 
\frac{ 
\left\langle E^{\,}_{n}\left| 
\frac{
\partial\mathcal{H}
     }
     {
\partial\varphi^{\,}_2
     }
\right|E^{\,}_{n'}\right\rangle 
\left\langle E^{\,}_{n'}\left| 
\frac{
\partial\mathcal{H}
     }
     {
\partial\varphi^{\,}_1
     } 
\right|E^{\,}_{n}\right\rangle
     }
     {
(E^{\,}_{n}-E^{\,}_{n'})^{2}
     }\,,
\end{split}
\label{eq:sigmaxy}
\end{equation}
where $q$ is the number of degenerate exact many-body ground states 
$\ket{E^{\,}_{n}}$ with the exact many-body energies $E^{\,}_{n}$, 
$\ket{E^{\,}_{n'}}$ denotes exact many-body excited states with 
the exact many-body energies $E^{\,}_{n'}$. 
The integrand is the Berry curvature of the exact many-body 
ground state $\ket{E^{\,}_{n}}$ and, 
even though it is not a perfectly flat function of 
$\varphi^{\,}_1$ 
and 
$\varphi^{\,}_2$, its integral over all 
$\bm{\varphi}\in[0,2\pi)^{2}$ 
values is expected to be quantized.

\section{Measuring band occupation}
In the main text, we defined the contribution of
the upper band to the FCI states as the overlap 
\begin{equation}
n^{\,}_+:=
\left\langle E^{\,}_{0}\left| 
\sum_{\bm{k}} 
\hat{n}^{\,}_{\bm{k},+} 
\right|E^{\,}_{0}\right\rangle,
\end{equation} 
where $|E^{\,}_{0}\rangle$ 
is any one of the states in the degenerate ground-state manifold 
and $\hat{n}^{\,}_{{\bm{k}},+}$ is the operator measuring density of particles 
with upper-band character at wave-number $\bm{k}$.

The eigendecomposition of Hamiltonian $\mathcal{H}$ 
can be written as 
$\mathcal{U}\,\mathcal{E}\,\mathcal{U}^{\dag}$, 
where $\mathcal{E}$ is a diagonal $2\times2$ matrix containing 
the single-particle eigenvalues $\varepsilon^{\,}_{+}$ and
$\varepsilon^{\,}_{-}$ 
and $\mathcal{U}$ is a unitary $2\times2$ matrix
containing the eigenstates of $\mathcal{H}$ as columns. 
We can now write the transformation from the sublattice to the band basis,
\begin{equation}
 \begin{pmatrix} 
\hat{c}^{\,}_{{\bm{k}},+} 
\\ 
\hat{c}^{\,}_{{\bm{k}},-} 
\end{pmatrix} 
= 
\mathcal{U}^{\dag} 
\begin{pmatrix} 
\hat{c}^{\,}_{{\bm{k}},A} 
\\ 
\hat{c}^{\,}_{{\bm{k}},B} 
\end{pmatrix} 
= 
\begin{pmatrix} 
u^*_{A,+}\,
\hat{c}^{\,}_{{\bm{k}},A} 
+ 
u^*_{B,+}\, 
\hat{c}^{\,}_{{\bm{k}},B} 
\\ 
u^*_{A,-}\, 
\hat{c}^{\,}_{{\bm{k}},A} 
+ 
u^*_{B,-}\, 
\hat{c}^{\,}_{{\bm{k}},B} 
\end{pmatrix}\,,
\end{equation}
where $u$ are the entries of $\mathcal{U}$. 
The density $\hat{n}^{\,}_{{\bm{k}},+}$ 
can therefore be written as
\begin{align}
\hat{n}^{\,}_{{\bm{k}},+} 
=& 
|u^{\,}_{A,+}|^2\,
\hat{n}^{\,}_{{\bm{k}},A} 
+ 
|u^{\,}_{B,+}|^2\,
\hat{n}^{\,}_{{\bm{k}},B} 
\nonumber\\
&
+ 
u^{\,}_{B,+}\, 
u^*_{A,+}\, 
\hat{c}^{\dag}_{{\bm{k}},B}\, 
\hat{c}^{\,}_{{\bm{k}},A} 
+ 
u^{\,}_{A,+}\, 
u^*_{B,+}\, 
\hat{c}^{\dag}_{{\bm{k}},A}\, 
\hat{c}^{\,}_{{\bm{k}},B}\,.
\end{align}
The exact many-body ground-state expectation value of any one of
$\hat{n}^{\,}_{{\bm{k}},A}$, 
$\hat{n}^{\,}_{{\bm{k}},B}$,
$c^{\dag}_{{\bm{k}},B}\,\hat{c}^{\,}_{{\bm{k}},A}$ 
and 
$\hat{c}^{\dag}_{{\bm{k}},A}\,\hat{c}^{\,}_{{\bm{k}},B}$ 
can be evaluated numerically on finite clusters for any value of $V$, 
using Lanczos exact diagonalization~\cite{Lanczos1950,Lanczos1952}.

\end{document}